\documentclass[prd,aps,floats,preprint,superscriptaddress,nofootinbib,tightenlines]{revtex4}
\usepackage{epsfig}
\usepackage{amsmath}
\def\euv{ \varepsilon_{\text{UV}} }
\def\eir{ \varepsilon_{\text{IR}} }
\def\lnQ{\ln\left(\frac{Q^2}{\mu^2}\right)}
\def\llnQ{\ln^2\left(\frac{Q^2}{\mu^2}\right)}
\def\e{\varepsilon}
\def\ol{\ln {\overline N}}
\def\dd{\delta(1-x)}

\def\Dsl{\hbox{/\kern-.6000em D}} 

\def\dsl{\,\raise.15ex\hbox{/}\mkern-13.5mu D}

\def\ltap{\ \raise.3ex\hbox{$<$\kern-.75em\lower1ex\hbox{$\sim$}}\ }
\def\gtap{\ \raise.3ex\hbox{$>$\kern-.75em\lower1ex\hbox{$\sim$}}\ }
\def\OMIT#1{}

\def\asf{\frac{\alpha_s}{4\pi}C_F}

\def\nsl{n\!\!\!\slash}

\def\OMIT#1{}

\newcommand{\nn}{\nonumber} 

\newcommand{\bn}{{\bar n}}
\newcommand{\bea}{\begin{eqnarray}}
\newcommand{\eea}{\end{eqnarray}}
\newcommand{\nb}{\bar n}

\begin{document}
\title{On The Equivalence of Soft and Zero-Bin Subtractions}

\author{Ahmad Idilbi}
\email{idilbi@phy.duke.edu}
\affiliation{Department of Physics, Duke University, Durham NC 27708, USA}

\author{Thomas Mehen}
\email{mehen@phy.duke.edu}
\affiliation{Department of Physics, Duke University, Durham NC 27708, USA}
\affiliation{Jefferson Laboratory, 12000 Jefferson Ave., Newport News VA 23606, USA}
\date{\today}
\vspace{0.5in}
\begin{abstract}

Zero-bin  subtractions are required to avoid double counting soft
contributions in collinear loop integrals in  Soft-Collinear Effective Theory
(SCET). In traditional approaches to factorization, double counting is avoided
by dividing  jet functions  by matrix elements of soft Wilson lines. In
this paper, we compare the two approaches to double counting, studying the quark form factor
and deep inelastic scattering (DIS) as $x_B \to 1$ as examples.  We explain how the zero-bin 
subtractions in SCET are required to reproduce the well-established factorization theorem for
DIS as $x_B \to 1$. We study one-loop virtual contributions to the quark form factor and 
real gluon emission diagrams in DIS.  The two approaches to  double counting
are equivalent if dimensional regularization (DR) is used to
regulate infrared (IR) divergences.  We discuss in detail ambiguities in the
calculation of one-loop scaleless integrals in DR in SCET and perturbative QCD. 
We also demonstrate a nontrivial check of the equivalence of the zero-bin subtraction and
the soft Wilson line subtraction in the virtual two-loop Abelian contributions to the quark form factor.

\end{abstract}

\maketitle

\section{Introduction}
High-energy processes with hadrons in  either the initial
or final state are amenable to perturbative QCD (pQCD) treatment
as long as there is a hard scale, $Q$,  much larger than
$\Lambda_{\rm QCD}$. The essential idea of pQCD factorization \cite{Collins:1989gx} is to separate the
different relevant scales into a well-defined quantities that capture
the physics at the given scale. By doing so one is then able to
distinguish the perturbative short-distance contribution from the
nonperturbative long-distance one.

In order to properly establish such
factorization of scales it is clear that one has to avoid double counting among the
 functions that appear in the factorized physical quantity. 
In general these nonperturbative functions receive leading contributions (in powers of
$\Lambda_{\rm QCD}^2/Q^2$) from specific regions in momentum space. For example, the contribution to a jet 
function may be dominated by partons with momenta whose components scale as
\bea \label{scaling}
(p^+,p^-,p^\perp) \sim Q(1,\lambda^2, \lambda) \, , 
\eea
where $\lambda$ characterizes the off-shellness of the partons in the jet function,
and $\lambda^2 \sim Q\Lambda_{\rm QCD}$ or $\Lambda_{\rm QCD}^2$, depending on the process considered.
In principle, one could restrict all momenta appearing in loop calculations 
to satisfy the scaling in Eq.~(\ref{scaling}), but this is impractical. Instead one typically integrates
over all momentum space in loop integrations, making approximations appropriate for the assumed scaling.
This  practice is well-known in, for example,  the method of regions~\cite{Beneke:1997zp}, where 
approximations appropriate to a specific momentum region are performed at the level of the
integrand, but the integrals are over all momentum space. In  cases where  contributions from 
other momentum regions are not sub-leading, one encounters double counting.

Consider, for example, deep inelastic scattering (DIS) in the end point region, $x_B \rightarrow 1$, where
$x_B$ is the well-known Bjorken variable. In this region, Sterman \cite{Sterman:1986aj} has shown that to all
orders in perturbation theory one can (re-)factorize the non-singlet DIS structure function in moment space
according to the  formulae
\bea\label{xft}
    F_{2,N}(Q^2) = H  \times \phi_N \times S_N \times  J_N\, ,
\eea
Here $F_{2,N}(Q^2)$ is the $N$th moment of the structure function and $x_B \to 1$ implies $N \gg 1$.
$H$ is the hard contribution, $\phi_N$ is the $N$th moment of a  modified  
parton distribution function (PDF),   $J_N$ is the $N$th moment of an outgoing jet function,
and $S_N$ is the $N$th moment of a soft factor representing the emission of 
soft gluons. Explicit and gauge invariant expressions for the PDF, jet, and soft functions will be given below. 
In the factorization theorem of Ref.~\cite{Sterman:1986aj}, the jet, soft, and parton distribution functions 
are not defined in a gauge invariant way, and through a judicious choice of gauge one can eliminate 
double counting. However when one defines the same quantities using Wilson lines to make them gauge
invariant, the factorization theorem breaks down and has to be modified to take into account the
double counting of soft momentum modes in the soft factor, $S_N$, and in the collinear matrix elements, $J_N$ and
 $\phi_N$. The Wilson lines in the matrix elements are the cause of the double counting problems.
 In Ref.~\cite{Akhoury:1998gs}, it was proposed that for gauge invariant quantities the 
 factorization theorem should take the form
\begin{equation}
  F_{2,N}(Q^2) = H  \left(\frac{\phi_N }{S_N }\right) S_N 
  \left(\frac{J_N }{S_N }\right)\, .
\label{dc}
 \end{equation} 
A similar factorization theorem with a soft factor subtracted from the transverse momentum dependent PDF and jet function was also
proposed by Ji, {\it et.\ al.} \cite{Ji:2004wu} for semi-inclusive DIS. Ref.~\cite{Ji:2004wu} checked the theorem
explicitly at one-loop and gave arguments for the theorem to hold to all orders in perturbation theory. Another well studied
quantity in pQCD is the quark form factor and its factorization into jets and soft contributions. The double counting problem for
this quantity is studied in  Refs.~\cite{Collins:1989bt,Collins:1999dz}, where it is argued that the same soft factor has to be
subtracted from each one of the two collinear jets, in a manner similar to Eq.~(\ref{dc}), to avoid double counting.

Recently, the problem of factorization has been addressed in an effective field theory 
approach using Soft Collinear Effective Theory (SCET)~\cite{Bauer:2000yr,Bauer:2001yt}. In this approach to 
factorization, QCD is matched onto SCET, a theory containing collinear and soft modes whose 
momenta scale as \footnote{In Ref.~\cite{Bauer:2000yr,Bauer:2001yt}, modes with all momentum components  
$O(Q\lambda)$ are called soft, while modes with all momentum components $O(Q\lambda^2)$ are 
called ultra-soft, or usoft. The $O(Q \lambda)$ modes play no role in this paper, so
we will neglect them, and we will refer to $O(Q \lambda^2)$ modes as soft rather than ultra-soft.}
\bea\label{cs}
{\rm collinear} &\sim& Q(1,\lambda^2,\lambda) \nn \\
{\rm soft} &\sim& Q(\lambda^2,\lambda^2,\lambda^2) \, .
\eea
After matching QCD onto SCET, a field redefinition~\cite{Bauer:2001yt} can be used to decouple the soft and
collinear modes at the level of the Lagrangian. In SCET, jet functions are matrix elements
of collinear fields and soft functions are matrix elements of soft fields. Note that in this paper
the terms soft and collinear are defined by Eq.~(\ref{cs}). For example, if a gluon whose
momentum components all scale as $O(Q\lambda^2)$ is emitted from a jet with small angle we still
refer to it as a soft gluon.

The subject of double counting has gained renewed interest in the context of 
SCET due to the work of Manohar and Stewart~\cite{Manohar:2006nz}. In SCET, power counting 
is made manifest using the label formalism~\cite{Georgi:1990um}. Let $\phi(x)$ denote a generic 
full theory field, the corresponding collinear field in SCET, $\phi_{\hat p}(x)$, is defined by
\bea
\phi(x) = \sum_{\hat p \neq 0} e^{-i \hat p \cdot x} \phi_{\hat p}(x) \, .
\eea
The $\hat p$ are $O(Q,Q\lambda)$
label momenta which correspond to the large parts of the collinear momenta,
while derivatives on $\phi_{\hat p}(x)$ give the $O(Q\lambda^2)$ residual 
momenta. Loop integrations involve both a sum over labels
and an integral
over the residual momentum which in practical calculations are combined into 
an integral over all momentum space,
\bea 
\sum_{\tilde p} \int d^d k  \longrightarrow \int d^d l \, .
\eea
But this raises the issue of double counting, as the integral over all momentum space
includes the region  in which the label momentum vanishes -
the ``zero-bin''  - which is already included in the soft sector of the theory.
To correctly calculate the collinear graphs, one must subtract the zero-bin contributions.
The zero-bin subtraction is necessary for the proper interpretation 
of $1/\epsilon$ poles in SCET loops~\cite{Manohar:2006nz}, and also necessary so that 
collinear plus soft real gluon emission graphs correctly reproduce real gluon emission QCD diagrams in the 
appropriate regions of phase space~\cite{Manohar:2006nz,Fleming:2006cd}.

The zero-bin subtraction in SCET is clearly related to the soft subtraction in Eq.~(\ref{dc}). In calculating the zero-bin
contribution to a particular Feynman graph, one must apply SCET power counting.  In the minimal zero-bin subtraction scheme of
Ref.~\cite{Manohar:2006nz} the zero-bin contribution is only removed if its contribution scales as $\lambda^0$, i.e., is
leading order in   SCET power counting. Only the zero-bin of the $O(Q \lambda^2)$ component of the gluon couples to collinear
fields at $O(\lambda^0)$  so in this scheme collinear couplings to other zero-bin modes can be neglected. Exploiting this
fact, Ref.~\cite{Lee:2006nr} shows that the difference between  calculating collinear matrix elements naively and including
the zero-bin subtraction amounts to using two different collinear Lagrangians that are related by a field redefinition of the
collinear fields~\cite{Lee:2006nr}. This  field redefinition is similar to the field redefinition that is used to  decouple
soft modes from collinear modes to prove  factorization in SCET. The field redefinition can be used to decouple the zero-bin mode from
collinear modes at the price of inserting zero-bin Wilson lines into the collinear matrix elements. Thus, the naively
evaluated collinear  matrix element factorizes into a properly evaluated collinear matrix   and a matrix element of zero-bin
Wilson lines, establishing the equivalence of zero-bin subtractions and soft Wilson line subtractions at lowest order in
$\lambda$. At higher orders in $\lambda$,  zero-bin modes of other components of the gluon field are important so it seems
that the equivalence only holds to  lowest order in $\lambda$.

The purpose of this paper is to study the equivalence between zero-bin
subtractions and the soft Wilson line subtractions in
more detail. We focus on the on-shell quark form factor and the factorization theorem for DIS as $x_B \to 1$. One important
issue is the choice of IR regulator in the calculation of the zero-bin contribution. The field redefinition which decouples
the soft or zero-bin modes and collinear modes will only leave on-shell matrix elements invariant, and therefore
off-shellness is not a suitable regulator~\cite{Bauer:2003td}. Soft Wilson line subtractions and zero-bin subtractions are not
equivalent if offshellness is used to regulate the IR divergences.

In our calculation of the quark form factor, we will use DR to regulate both ultraviolet (UV) and IR divergences.
The individual one-loop 
soft diagrams, the collinear diagrams and their zero-bin subtractions are all ill-defined, and only
their sum yields a well-defined result which reproduces the IR divergences of QCD. 
Fortunately, in this case the equivalence of the soft   and the zero-bin subtractions
can be easily established at the level of the integrands.
At higher orders the zero-bin subtraction involves an iterative procedure which 
requires the subtraction of graphs in which some lines have collinear momentum and some  
have soft momentum, so demonstrating the equivalence of soft and zero-bin
subtractions at higher orders is somewhat subtle. For example, one has to be careful about the
order in which limits are taken to correctly compute the zero-bin subtraction at two loops.
Another issue is that the equivalence of zero-bin and soft subtractions does not hold at the level 
of individual Feynman diagrams, but only for the sum of all diagrams at a given order.
In some cases this requires a cancellation between the zero-bins that appear in different diagrams.  In this paper,
we explore these issues by calculating the zero-bin subtractions in the two-loop 
abelian diagrams contributing to the quark form factor. We also show that they are equivalent to soft subtraction. 

Another important point of this paper is to emphasize the importance of the zero-bin subtractions in
the collinear matrix elements that appear in the SCET derivation of the endpoint factorization in DIS as $x_B \to 1$. 
We will assume a factorization theorem of the form of Eq.~(\ref{xft}) where the definitions 
of $S_N$, $\phi_N$, and $J_N$ in terms of SCET fields will be given below. We will show that  
the one-loop zero-bin contributions to $\phi_N$ and $J_N$ are equal to $S_N$, which confirms Eq.~(\ref{dc}).
We will also show that at one-loop Eq.~(\ref{dc}) reproduces the large $N$ limit of the full QCD calculation
of DIS as $x_B \to 1$. In these calculations we again use DR to regulate the IR 
divergences. Our  results are consistent with  those
in Ref.~\cite{Chen:2006vd}, however, in our treatment all Wilson lines are defined on the light-cone.
For other SCET treatments of DIS in the $x_B \to 1$ limit, see Refs.~\cite{Manohar:2003vb,Becher:2006mr,Chay:2005rz}.

This paper is organized as follows. In section II, we discuss the space-like quark form factor in SCET and present one-loop results
for the various contributions. Once zero-bin contributions are included it is straightforward to verify that SCET reproduces the IR
divergences of QCD.  In section III, we give our analysis of DIS at $x \to 1$ in SCET. In section IV, we study the
zero-bin subtractions for two-loop virtual abelian contributions to the quark form factor and demonstrate agreement with 
the soft Wilson line subtractions. In section V we conclude.

\section{Zero-Bin at One-Loop: Quark Form Factor}

In this section we calculate the quark form factor to one loop using DR 
to regulate both the UV and IR. It is straightforward 
to see that the zero-bin subtractions are reproduced by the soft subtractions 
for this choice of regulator.
The quark form factor may be the simplest quantity to analyze in
SCET, however it serves well to establish some of the relevant
issues. We take the incoming quark to be moving along the $+z$ direction
so its four-momentum is $p_{1}^{\mu}=(p_1^+,p_1^-,p_1^\perp)=(\frac{Q}{\sqrt2},0,0)=\frac{Q}{\sqrt2} n^\mu$.
The incoming quark scatters off a space-like photon whose invariant mass
is $q^2\equiv-Q^2 <0$. The final state quark is moving in the $-z$ 
direction so its momentum is $p_{2}^{\mu}=(p_2^+,p_2^-,p_2^\perp)=
(0,\frac{Q}{\sqrt2},0) = \frac{Q}{\sqrt2} \bn^\mu$.

The full QCD electromagnetic current, $j^{\mu}={\bar \psi}(x)\gamma^{\mu}\psi(x)$,
is matched onto an effective one,
\begin{eqnarray}
j^{\mu}&=&C(Q^2)j^{\mu}_{eff}= C(Q^2)\ \bar{\xi}_{\bar n} W_{\bar n}
\gamma^{\mu}W_n^\dagger \xi_n \nonumber \\
&=&C(Q^2)\ \bar{\xi}^{(0)}_{{\bar n}} W^{(0)}_{{\bar
n}} Y^{\dagger}_{{\bar n}} \gamma^{\mu} Y_n W^{(0)\dagger}_n\xi^{(0)}_n \, .
 \label{current}
\end{eqnarray}
The manipulations performed in Eq.~(\ref{current}) are 
standard in the SCET formalism. The form of the current is fixed by collinear
gauge invariance. The second line is a consequence of making the
field redefinition of the collinear quark and gluon fields that decouples soft gluons from 
the collinear  sector of the leading order SCET lagrangian. The collinear Wilson line, $W_n$,
and the soft Wilson line, $Y_n$, are given in momentum space by
\begin{eqnarray}
\label{wil}
W_n=\sum_{m=0}^{\infty}\sum_{{\rm perms}}\frac{(-g)^m}{m!}\frac{\bar
n \cdot A_{n,q_1}\cdots \bar n \cdot A_{n,q_m}}{\bar n \cdot q_1
\bar n \cdot (q_1+q_2)\cdots \bar n \cdot (\sum_{i=1}^{m} q_i)} \ ,
\end{eqnarray}
 and
\begin{eqnarray}
Y_n=\sum_{m=0}^{\infty}\sum_{{\rm perms}}\frac{(-g)^m}{m!}\frac
{n\cdot A_s\cdots n\cdot A_s}{n\cdot q_1 n\cdot(q_1+q_2)\cdots
n\cdot (\sum_{i=1}^m q_i)} \ .
\end{eqnarray}
$A_{n,q_i}$ is a collinear gluon field and all $\bar n\cdot q_i$
scale as $Q$. $A_s$ stands for soft gluon field and $n\cdot q_i$
scales as $Q\lambda^2$.

The matrix element of $j_{eff}^\mu$ factorizes   into 
\begin{eqnarray} 
\langle q(p_2)\vert j^{\mu} \vert q(p_1)\rangle=C(Q^2)\gamma^{\mu}[J_{out} \times \gamma^{\mu} \times J_{in} \times S] \, ,
\end{eqnarray}
where the jet and soft functions are defined by
\begin{eqnarray}J_{in}=\langle 0\vert
W^{(0)\dagger}_n \xi^{(0)}_n \vert q(p_1)\rangle, \,\,\,\,J_{out}=\langle q(p_2)\vert {\bar \xi^{(0)}}_{{\bar n}} W^{(0)}_{{\bar n}}\vert 0 \rangle,
\,\,\,\, S=\langle 0\vert Y^{\dagger}_{\bar n}Y_n \vert 0 \rangle \,\, . 
\end{eqnarray} 
In the jet functions the fields are collinear so only fields with nonvanishing labels appear. 
As argued by Lee and Sterman~\cite{Lee:2006nr}, only  the zero-bin of the $n\cdot A_{n,q}$ 
component of the collinear gluon couples to $\xi^{(0)}_n$ at leading order in $\lambda$.
Furthermore, the Wilson line, $W_n^{(0)\dagger}$, which has no zero-bin gluon is related to one with 
a zero-bin by multiplication by a Wilson line constructed from  the zero-bin  component of the
$\bn\cdot A_{n,q}$ gluon. 
Thus the collinear matrix element is equal to the naively evaluated matrix 
element  divided by a matrix element of soft Wilson lines:
\begin{eqnarray} \label{fun} \langle 0\vert  W^{(0)\dagger}_n \xi^{(0)}_n\vert q(p_1)\rangle = 
\frac{\langle 0\vert {\hat W}^{\dagger}_n {\hat \xi}_n \vert q(p_1)\rangle}
{\langle 0\vert Y^{\dagger}_{\bar n}Y_n \vert 0 \rangle} \,\, ,
\end{eqnarray}
where 
\begin{eqnarray}
 Y_n(x)=P \mathrm {exp} \left[ ig_s\int_{-\infty}^{0}ds n\cdot A_{n,0}(ns+x)\right]\,\, ,
\end{eqnarray}
$A_{n,0}$ is the zero-bin of the collinear gluon, and our definition of the covariant derivative is $D^{\mu}=\partial^{\mu}-ig_sA^{\mu}$.
We will use the same notation for the soft Wilson line and the zero-bin Wilson line because
the matrix elements are identical and it  is always clear from context which is relevant.
In Eq.~(\ref{fun}) the hats on the fields signify that  $\hat W_n^\dagger$ contains the zero-bin gluon and that
the Lagrangian includes the coupling of  $\hat \xi_n$ to zero-bin gluons, so this matrix element corresponds to the 
naive evaluation of the collinear matrix element.  In our notation 
$\langle 0\vert W^{(0)\dagger}_n \xi^{(0)}_n \vert q(p_1)\rangle$ corresponds to 
$\langle 0\vert W'^{\dagger}_n {\xi'}_n \vert q(p_1)\rangle$ of Ref.~\cite{Lee:2006nr},
$\langle 0\vert {\hat W}^{\dagger}_n {\hat \xi}_n \vert q(p_1)\rangle$ corresponds to 
$\langle 0\vert  W''^{\dagger}_n \xi''_n \vert q(p_1)\rangle$, $Y_n$ corresponds to $U_n^{\dagger}$,
and $Y_{\bar n}$ corresponds to $\Omega_n$.

\begin{figure}[t]
 \begin{center}
 \includegraphics[width=3.0in]{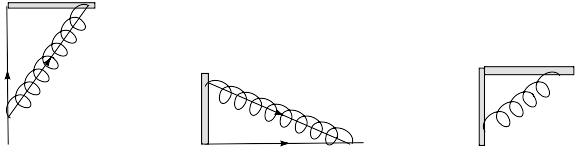}
 \end{center}
 \vskip -0.7cm \caption{The quark form factor in SCET. In the first two 
 graphs  the thick line attached  to the fermion line represents a collinear 
 Wilson line, while in the last graph both thick lines are soft Wilson  lines.}
  \label{fullcurrent}
 \end{figure}

At one-loop the Feynman diagrams that contribute to the
collinear $n$-jet, collinear $\bar n$-jet and the soft factor are
given in Fig.~\ref{fullcurrent}.  All these diagrams  are scaleless in DR and
therefore all integrals will vanish when one sets $\euv=\eir$. 
Using SCET Feynman rules to calculate the $n$-collinear diagram and 
naively integrating  over all momentum space, the result is
\begin{equation}
\label{coln1}
 I_n=-2ig_s^2C_F(\mu^2)^{\varepsilon}\int
\frac{d^dk}{(2\pi)^d}\frac{{\tilde p}-k^+}{[-2{\tilde p}k^-+k^2+i0][-k^++i0][k^2+i0]}
\ ,
\end{equation}
where ${\tilde p}\equiv p_1^+=p_2^-=\frac{Q}{\sqrt 2}$. Note that the collinear graph
is not    $I_n$ but rather $I_n - I_{n,0}$, where $I_{n,0}$ is the zero-bin contribution. We  refer 
to $I_n$ as the naive $n$-collinear contribution.
The naive ${\bar
n}$-collinear contribution is similar,
\begin{equation}
\label{coln2}
 I_{\bar n}=-2ig_s^2C_F(\mu^2)^{\varepsilon}\int
\frac{d^dk}{(2\pi)^d}\frac{{\tilde p}-k^-}{[-2{\tilde p}k^++k^2+i0][-k^-+i0][k^2+i0]}
\ ,
\end{equation}
and the soft contribution is
\begin{equation}
\label{is}
I_s=-ig_s^2C_F(\mu^2)^{\varepsilon}\int
\frac{d^dk}{(2\pi)^d}\frac{1}{(-k^++i0)(-k^-+i0)(k^2+i0)} \ .
\end{equation}
Let us now compute the zero-bin subtraction for $I_n$. It is easy to check that 
the zero-bin associated with the virtual fermion line 
is subleading in $\lambda$. The only zero-bin needed is associated with the virtual gluon.
In this zero-bin region, $k^+$ 
scales as $Q\lambda^2$ instead of $Q$ so the numerator is simply 
$\tilde p$. Furthermore,  $k^2$ scales
$Q^2\lambda^4$ instead of $Q^2\lambda^2$ thus it should be dropped
relative to ${\tilde p}k^-$. Similar arguments hold for $I_{\bar n}$. The
result is simply that
\begin{eqnarray}
I_{n,0}&=&I_{\bar {n},0}=-ig_s^2C_F(\mu^2)^{\e}\int
\frac{d^dk}{(2\pi)^d}\frac{1}{(-k^++i0)(-k^-+i0)(k^2+i0)}\, \nn \\
&=& I_s
\end{eqnarray}
The complete one-loop SCET contribution to the quark form factor is 
\bea\label{zbes}
I &=& (I_n-I_{n,0})+(I_{\bar n}-
I_{{\bar n},0})+I_s \nn \\
&=& I_n + I_\bn -I_s \, .
\eea
Note that the net effect of the zero-bin subtractions  at one-loop is to reverse the 
sign of the soft contributions.  
Thus the zero-bin subtraction and soft subtraction of Eq.~(\ref{dc}) yield the 
same results for the one-loop graphs when they are regulated in DR.  

One subtlety in Eq.~(\ref{zbes}) is that the integrals $I_n$, $I_\bn$ and 
$I_s$ are all ill-defined. However, in Eq.~(\ref{zbes}) the integrands can be rearranged 
so that $I$ is expressed in terms of integrals which are well-defined,
then it is straightforward to show that $I$ reproduces the IR of QCD. 
 Eq.~(\ref{zbes}) can be written as
\begin{eqnarray}
I&=&-2ig_s^2C_F(\mu^2)^{\e}\int
\frac{d^dk}{(2\pi)^d}\left\{
\frac{2{\tilde p}[{\tilde p}-k^+-k^-]}{[-2{\tilde p}k^++k^2+i0][-2{\tilde p}k^-+k^2+i0][k^2+i0]}\right.\nonumber\\
&&+\frac{2}{[-2 {\tilde p}k^++k^2+i0][-2{\tilde
p}k^-+k^2+i0]}  \nn \\
&&\left. -\frac{k^2}{2[-2 {\tilde p}k^++k^2+i0][-2{\tilde
p}k^-+k^2+i0][-k^++i0][-k^-+i0]}
\right \} \nn \\
&\equiv& I_1 +I_2+I_3
\,\, , \label{f/h}
\end{eqnarray}
where
\begin{eqnarray}
I_1&=&-2ig_s^2C_F(\mu^2)^{\e}\int
\frac{d^dk}{(2\pi)^d}\frac{2{\tilde p}(\tilde p - k^+-k^-)}{[-2{\tilde p}k^++k^2+i0][-2{\tilde p}k^-+k^2+i0][k^2+i0]}\nonumber\\
&=&-\frac{\alpha_s}{2\pi}C_F\left(\frac{4\pi\mu^2}{Q^2}\right)^{\e}\frac{\Gamma(1+\e_{\rm
IR})\Gamma^2(-\e_{\rm IR})}{\Gamma(2-2\e_{\rm IR})}\,\, , \\
&& \nn \\
I_2&=&-2ig_s^2C_F(\mu^2)^{\e}\int
\frac{d^dk}{(2\pi)^d}\frac{2}{[-2{\tilde p}k^++k^2+i0][-2{\tilde p}k^-+k^2+i0]}\nonumber\\
&&=+\frac{\alpha_s}{2\pi}C_F\left(\frac{4\pi\mu^2}{Q^2}\right)^{\e} 
\frac{2 \,\Gamma^2(1-\e_{\rm UV})\Gamma(\e_{\rm
UV})}{\Gamma(2-2\e_{\rm UV})}\,\, ,\\
&& \nn \\
I_3&=&-2ig_s^2C_F(\mu^2)^{\e}\int
\frac{d^dk}{(2\pi)^d}\frac{-k^2}{2[-2{\tilde p}k^++k^2+i0][-2{\tilde p}k^-+k^2+i0][-k^++i0][-k^-+i0]}\nonumber\\
&&=+\frac{\alpha_s}{2\pi}C_F\left(\frac{4\pi\mu^2}{Q^2}\right)^{\e}\frac{1}{\e_{\rm UV}^2}
\frac{\Gamma(1-\e_{\rm UV})^2\Gamma(1+\e_{\rm UV})}{\Gamma(1-2\e_{\rm
UV})}\,\,. \label{I4}
\end{eqnarray}
Before we combine these integrals we comment on the identification of UV and IR poles.
The integrals in $I_1$ and $I_2$ can be calculated
by combining the propagators using Feynman parameters and completing 
the square in the standard manner. $I_1$ is  clearly UV finite and IR divergent
by power counting, and $I_2$ is clearly UV divergent and IR finite by power
counting. In $I_3$, if we rescale the loop momenta homogeneously
we conclude that the integral is UV divergent and IR finite. One may worry if there are
IR divergences in $I_3$ that come from $k^+\to 0$ or  $k^- \to 0$ with other
components held fixed. In the Appendix, we describe a careful
evaluation of the integral using contour integration that  confirms that the poles are UV.

It is interesting that the scale $Q^2$ appears naturally in the evaluation of the integrals $I_i$.
The integrals $I_n$ and $I_\bn$ and $I_s$ are scaleless and cannot know 
about the scale $Q$ unless it is put in by hand. We will see this when we attempt to evaluate
these integrals individually later in this section. However, after rearranging the integrands,
$I$ is expressed in terms of integrals, $I_i$, $i = 1,2,3$, in which the scale $Q$ naturally 
appears. It is also important to notice that the individual
contributions, $I_i$, are free from mixed UV/IR poles thus there is no ambiguity in interpreting the
$\varepsilon$ in $(\mu^2/Q^2)^\varepsilon$.

The final result for $I$ is
\begin{eqnarray}
 I&=&
 \frac{\alpha_s}{4\pi}C_F\left\{\left(\frac{2}{\e^2_{\rm
 UV}}-\frac{2\ln \left(\frac{Q^2}{\mu^2}\right)-4}{\e_{\rm
 UV}}\right)-\left(\frac{2}{\e^2_{\rm
 IR}}-\frac{2\ln \left(\frac{Q^2}{\mu^2}\right)-4}{\e_{\rm
 IR}}\right)\right\}\,\,.
\label{aa}
 \end{eqnarray}
We must also include the factor of $\sqrt {Z_2}$ for each external leg, where
\begin{eqnarray}
Z_2=1-\frac{\alpha_s}{4\pi}C_F\left(\frac{1}{\euv}-\frac{1}{\eir}\right)\,\, .
\end{eqnarray}
The final result for the electromagnetic current in SCET to one loop is
\begin{eqnarray}
\label{disscet} \langle q(p_2)\vert j^{\mu}_{eff}\vert q(p_1)\rangle=\gamma^\mu\left[1+\frac{\alpha_s}{4\pi}C_F\left(\frac{2}{\varepsilon^2_{\rm
UV}}-\frac{2\ln\left(\frac{Q^2}{\mu^2}\right)-3}{\varepsilon_{\rm
UV}}-\frac{2}{\varepsilon_{\rm
IR}^2}+\frac{2\ln\left(\frac{Q^2}{\mu^2}\right)-3}{\varepsilon_{\rm
IR}}\right)\right] \, .
\end{eqnarray}
The UV poles are canceled by counterterm for the SCET effective current,
\bea
{\rm c.t.}=\asf \left[-\frac{2}{\euv^2}-\frac{3}{\euv}+\frac{2}{\euv}\lnQ \right]\,\,.
\eea
The IR poles
are exactly the same  as in the full QCD calculation, which is given by
\begin{equation}
\label{disfull} \langle q(p_2)\vert j^\mu\vert q(p_1)\rangle=
\gamma^\mu\left[1+\frac{\alpha_s}{4\pi}C_F\left(-\frac{2}{\varepsilon_{\rm
IR}^2}+\frac{2\ln\left(\frac{Q^2}{\mu^2}\right)-3}{\varepsilon_{\rm IR}}
-\ln^2\left(\frac{Q^2}{\mu^2}\right)+3\ln\left(\frac{Q^2}{\mu^2}\right)-8+\frac{\pi^2}{6}\right)\right]
\,\, .
\end{equation}
and the matching coefficient of the full QCD current onto the SCET one is just the finite part of Eq.~(\ref{disfull}):
\bea
C(Q^2/\mu^2)=1+\asf\left[-\llnQ+3\lnQ-8+\frac{\pi^2}{6}\right]\,\,,
\label{cc}
\eea
from  which the anomalous dimension of the effective current $\gamma_1$ is obtained:
\bea
 \frac{d\ln C}{d\ln \mu}=\gamma_1 \,\,,
\label{ano}
\eea
with
\bea
\gamma_1=\asf \left[-4\lnQ-6 \right]\,\,.
\eea

The above results for $C(Q^2/\mu^2)$ and $\gamma_1$ where first obtained in Ref.~\cite{Manohar:2003vb},
where offshellness was used to regulate the IR divergences.
Next we attempt to directly evaluate the the integrals $I_n$ and $I_s$.
It is possible to unambiguously determine the double poles in $\epsilon$
in these graphs but the single poles are ambiguous. Consider first the naive
collinear contribution $I_n$, which can be written as 
\begin{eqnarray}\label{evalin}
I_n&=&-2ig_s^2C_F(\mu^2)^{\varepsilon}\left[\int
\frac{d^dk}{(2\pi)^d}\frac{{\tilde p}}{[-2{\tilde p}k^-+k^2+i0][-k^++i0][k^2+i0]}\right.\nonumber\\
&&\left.+\int\frac{d^dk}{(2\pi)^d}\frac{1}{[-2{\tilde p}k^-+k^2+i0][k^2+i0]}\right]\,\,,
\end{eqnarray}
To calculate the first integral in Eq.~(\ref{evalin}), begin by performing the $k^+$ integral using contour
integration, then do the $d-2$ dimensional integral over the transverse momentum. One is left with an integral
over $k^-$ which is proportional to
\bea
\int dk^- (2 \tilde p k^-)^{-1-\epsilon} = Q^{-2 \epsilon} \left(\frac{1}{\e_{\rm UV}} - \frac{1}{\e_{\rm UV}}\right)
\eea
In this formula, we have rescaled $k^- \to \tilde p \, k^-$ then used a standard result in dimensional regularization.
The rescaling is required so that that the equation is dimensionally correct in $d$ dimensions but the dimensional
quantity  that is raised to $-2 \epsilon$ power is arbitrary and we have put the scale $Q$  into the integral by hand. 
The second integral in Eq.~(\ref{evalin}) can be easily calculated by combining the integrals
using standard Feynman parameterization and the result is
\bea
\mu^{2 \epsilon} \int \frac{d^dk}{(2\pi)^d k^4}=\frac{i}{16\pi^2}\left(\frac{ \mu^2}{Q^2}\right)^{\epsilon}
\left(\frac{1}{\euv}-\frac{1}{\eir}\right) \,\,,
\label{pam}
\eea
where again the scale $Q$ has been inserted to make the result sensible on dimensional grounds.
 
The final result is
\begin{eqnarray}
I_n=\frac{\alpha_s}{4\pi}C_F\left(\frac{\mu^2}{Q^2}\right)^{\varepsilon}\left[\frac{1}{\eir}\left(\frac{2}{\euv}-\frac{2}{\eir}\right)+\left(\frac
{2}{\euv}-\frac{2}{\eir}\right)\right]\,\,.
\label{In}
\end{eqnarray}
There are two sources of ambiguity in the evaluation of this integral: the scale $Q$ has been inserted by hand, as
discussed earlier, and furthermore, the expansion in $\epsilon$ is ambiguous because of the mixed $1/(\epsilon_{\rm UV}
\epsilon_{\rm IR})$ pole. Because of this, our result for $I_n$ (and $I_\bn$)
should really be regarded as a prescription for  defining the integral. A similar situation arises in the 
evaluation of $I_s$, as we will see below. 

For the soft diagram, the contour integration fails and one has
to perform the integral differently. Inserting by hand a
scale in the integral, we rewrite $I_s$ in the form
\begin{eqnarray}
I_s=-2ig_s^2C_F(\mu^2)^{\varepsilon}\frac{Q^2}{2}\int\frac{d^dk}{(2\pi)^d}\frac{1}{(p_1\cdot
k)(p_2\cdot k)k^2}\,\, ,
\end{eqnarray}
where, as before, $p_1$ $(p_2)$ stands for the momentum of the
incoming (outgoing) parton with $p_1^+ (p_2^-)=\frac{Q}{\sqrt 2}$
and $p_i^2=0$. Using the following identity
\begin{eqnarray}
\frac{1}{p_i\cdot k}=\frac{2}{(p_i+k)^2}\left[1+\frac{k^2}{2p_i\cdot
k}\right]\,\, ,~~~i=1,2\,\, ,
\end{eqnarray}
the integral becomes
\begin{eqnarray}
I_s=-2ig_s^2C_F(\mu^2)^{\varepsilon}Q^2\times
\left[I_{s,1}+I_{s,2}+I_{s,3}\right]\,\, ,
\label{pre}
\end{eqnarray}
with
\begin{eqnarray}
I_{s,1}=\int
\frac{d^dk}{(2\pi)^d}\frac{1}{(p_1+k)^2(p_2+k)^2k^2}\,\, ,
\end{eqnarray}
\begin{eqnarray}
I_{s,2}=\int
\frac{d^dk}{(2\pi)^d}\frac{1}{(p_1+k)^2(p_2+k)^2}\left[\frac{1}{2p_1\cdot
k}+\frac{1}{p_2\cdot k}\right]\,\, ,
\end{eqnarray}
and
\begin{eqnarray}
I_{s,3}=\int \frac{d^dk}{(2\pi)^d}\frac{k^2}{(p_1+k)^2(2p_1\cdot
k)(p_2+k)^2(p_2\cdot k)}\,\, .
\end{eqnarray}
The calculation of $I_{s,1}$ is straightforward. Multiplying the result with the pre-factor in Eq.~(\ref{pre}) and denoting the result
${\tilde I}_{s,1}$ one obtains
\begin{eqnarray}
{\tilde I}_{s,1}=\frac{\alpha_s}{4\pi}C_F\left(\frac{ \mu^2}{Q^2}\right)^{\varepsilon}\left[-\frac{2}{\varepsilon^2_{\rm IR}}
+\frac{\pi^2}{6}\right]\,\, .
\end{eqnarray}
 For $I_{s,2}$ and $I_{s,3}$, the first step is to combine
 \bea
 \frac{1}{2 p_i \cdot k} \frac{1}{( p_i + k)^2} = \frac{1}{2 p_i \cdot k} \frac{1}{k^2 + 2 p_i \cdot k} \, , \nn
 \eea
using the identity
 \begin{eqnarray}
 \frac{1}{b(a+b)}=\int_1^{\infty}d\lambda \frac{1}{[a+\lambda
 b]^2}\,\, .
 \end{eqnarray}
 After this step the evaluation of the integral is straightforward and we find
 \begin{eqnarray}
 {\tilde I}_{s,2}=\frac{\alpha_s}{4\pi}C_F
 \left(\frac{ \mu^2}{Q^2}\right)^{\varepsilon}\left[\frac{4}{\euv
 \eir}- \frac{\pi^2}{3}\right]\,\, ,
 \end{eqnarray}
 and
 \begin{eqnarray}
 {\tilde
 I}_{s,3}=\frac{\alpha_s}{4\pi}C_F\left(\frac{ \mu^2}{Q^2}\right)^{\varepsilon}\left[-\frac{2}{\varepsilon^2_{\rm
 UV}}+ \frac{\pi^2}{6}\right]\,\, .
 \end{eqnarray}
 The combined result gives
 \begin{eqnarray}
 I_s=\frac{\alpha_s}{4\pi}C_F
 \left(\frac{ \mu^2}{Q^2}\right)^{\varepsilon}(-2)\left[\frac{1}{\euv}-\frac{1}{\eir}\right]^2\,\,.
\label{sam}
 \end{eqnarray}
Again, $I_s$ is ambiguous because the scale that compensates $\mu$ is arbitrary and because of mixed 
$1/(\e_{\rm UV}\e_{\rm IR})$ poles. 
When our results for $I_n$, $I_\bn$ and $I_s$ are combined according to Eq.~(\ref{zbes}), the mixed 
$1/(\e_{\rm UV} \e_{\rm IR})$ poles in $I$ cancel. We can then separate the UV and IR divergent
terms, expand $( \mu^2/Q^2)^{\e}$ in each term and recover Eq.~(\ref{aa}). The prescriptions for 
defining $I_n$ and $I_{\bn}$ and $I_s$ are thus justified {\it a postieri} by the requirement
that SCET reproduce the IR divergences of QCD.

Our main result for this section is that the soft subtraction in Eq.~(\ref{dc}) gives the same
result as the zero-bin subtraction when the one-loop graphs for the jet and soft 
functions are evaluated using  DR to regulate the IR as well as UV. 
While the one-loop integrals in
the evaluation of  the collinear and soft functions are ill-defined, it is possible 
to see the equivalence of zero-bin and soft subtraction at the level of the integrands. The sum of one
loop collinear and soft graphs is well-defined and reproduces the IR divergences of QCD. We gave 
a prescription for evaluating these integrals which reproduces these results. 

\section{DIS as $x \to 1$: Soft Subtraction}

In this section we extend the analysis of the quark form factor  and zero-bin subtractions presented in Section II to the DIS
non-singlet structure function $F_2(x,Q^2)$ in the threshold region, $x \to 1$. We follow the notation of Ref.~\cite{alta}
with $x=Q^2/2p_1\cdot q$ and $p_1$ is the momentum of the incoming parton. We define all quantities to be gauge invariant. All
Wilson lines are defined on the light-cone and as before we regularize both IR and UV divergences  in pure DR. It will be
shown at  $O(\alpha_s)$ that the zero-bin contributions exist in Feynman diagrams with real gluon emission that contribute to
the naive collinear matrix elements. For DIS these are the well-known jet function and  PDF to be defined below. Moreover we
will see that the zero-bin is equivalent to the soft contribution in pure DR. When eliminating the double counting by
generalizing Eq.~(\ref{fun}) to take into account a product of two electromagnetic currents we recover, in the SCET formalism,
the factorization theorem given in Eq.~(\ref{dc}) that holds to all orders in perturbation theory.

The general strategy we advocate to eliminate double counting in SCET factorization formalism for a given physical quantity is
to first decouple the soft gluons from collinear SCET Lagrangian by performing field redefinition: $\xi \rightarrow \xi^{(0)}$
and $A^{\mu}\rightarrow A^{(0),\mu}$. In what follows, we drop the superscript $(0)$ with the understanding that decoupling
the soft gluons is already performed.  Then one defines \emph {naive} collinear matrix elements and includes in the SCET
Lagrangian the zero-bin fields. This enables us to perform perturbative calculations extended over all momentum space which 
include contributions from the soft momentum region. These contributions should then be eliminated by performing the zero-bin
subtractions. We will show at one-loop order that these zero-bin subtractions are equivalent to dividing by matrix elements of
soft Wilson lines, confirming Eq.~(\ref{dc}). The validity of this procedure can be justified to all orders in perturbation
theory and to lowest order in $\lambda$  following the arguments of Ref.~\cite{Lee:2006nr}.

\begin{figure}[h]
\begin{center}
\includegraphics{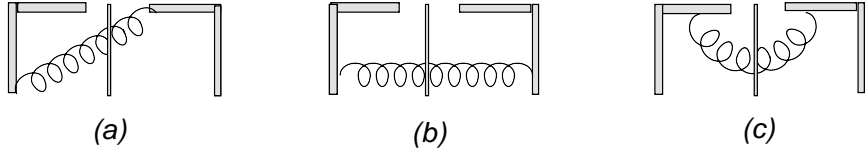}
\end{center}
\vskip -0.7cm \caption{Real gluon contribution to the soft factor.}
\label{SOFTR}
\end{figure}

Let us start by defining the soft factor $S$ which is given by,
 \begin{eqnarray}
  S(1-x)&=&\frac{{\tilde p}}{N_c}\int \frac{d\lambda}{2\pi} e^{i\lambda (1-x){\tilde p}}
    \langle 0|{\rm Tr}[Y_{n}(\lambda n)Y^{  \dagger}_{\bn}(\lambda \bn)
     \times Y^{ \dagger}_{\bn}(0)Y_n(0)]|0\rangle\,\,, \nonumber 
    \end{eqnarray}
where ${\tilde p}=Q/{\sqrt 2}$. The pre-factor is chosen to normalize
 the leading contribution to $\delta(1-x)$. The Feynman diagrams with real gluon emission that contribute to $S(1-x)$ are given in Fig.~\ref{SOFTR}.  
 Figs.~\ref{SOFTR}(b) and \ref{SOFTR}(c) are identically zero due to $n^2=\bn^2=0$ respectively. The contribution from  Fig.~\ref{SOFTR}(a) is
\bea
\label{ssr1} S^{(a)}(1-x)&=&2\alpha_s C_F (\mu^2)^{\e}{\tilde p}
\int \frac{d^d k}{(2\pi)^{d-2}} \frac{1}{k^+k^-}\delta(k^2)\delta(k^+-(1-x){\tilde p})\nonumber\\
&&=\frac{\alpha_s}{4\pi}C_F\left(\frac{\mu^2}{Q^2}\right)^\e\left[-\frac{1}{\eir}\delta(1-x)+D_0(x)\right]\left(\frac{2}{\euv}-\frac{2}{\eir}\right)\,\,,
\eea
and the $Q^2$-dependence is again put by hand as the integral is scaleless. We use $D_i(x)$ for the ``plus'' distributions:
\begin{eqnarray}
 D_i(x) \equiv \left
 [\frac{\ln^{i}(1-x)}{1-x}\right]_+\,\,\,\,\,i=0,1.
 \end{eqnarray}
In going from the first to second line in Eq.~(\ref{ssr1}) we have carried out the integral over $k^+$ first, 
then integrated over the transverse momentum,  and finally integrated $k^-$ from $0$ and $\infty$. 
 Taking the contribution from the mirror diagram
of (a) and adding  $2I_s\times \delta(1-x)$ ($I_s$ is given in Eq.~(\ref{sam})) to include the virtual contributions 
we get the soft factor to   $O(\alpha_s)$ in pure DR:
\bea
\label{ss1}
S(1-x)&=&\delta(1-x) \\
&&+\frac{\alpha_s}{4\pi}C_F\left(\frac{\mu^2}{Q^2}\right)^\e\times 
4\left[\left(-\frac{1}{\euv^2}+\frac{1}{\euv\eir}\right)\delta(1-x)+D_0(x)\left(\frac{1}{\euv}-\frac{1}{\eir}\right)\right]\,\,. \nn
\eea
The renormalized soft function to $O(\alpha_s)$ is 
\bea
\label{ssr}
S_R(1-x)=\delta(1-x)+\frac{\alpha_s}{4\pi}C_F\times (-4)\frac{D_0(x)}{\eir}\,\,.
\eea

\begin{figure}[h]

\begin{center}

\includegraphics{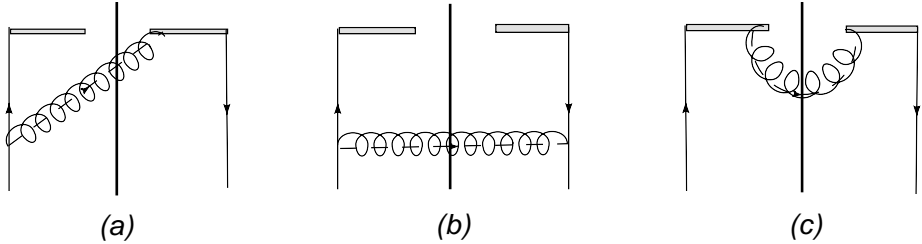}
\end{center}
\vskip -0.7cm \caption{Real gluon contribution to the PDF.}
\label{PDFR1}
\end{figure}

Now we consider the PDF. Our definition of the PDF is analogous to the standard one in QCD \cite{col} 
however it is expressed in terms of the naive collinear SCET fields,
\bea
{\hat \phi} (x)  = \frac{1}{2}\int \frac{d\lambda}{2\pi}
 e^{i\lambda x\!{\tilde p}} \langle P | \overline{{\hat \xi}}_n(\lambda \bn) {\hat W}_{
 n}(\infty,0;\lambda \bn)  \times
  \gamma^+ {\hat W}^\dagger_n(\infty,0;0){\hat \xi}_n (0)|P\rangle \, . \label{pdf}
\eea
The Feynman diagrams with real gluon emission that contribute to the partonic PDF are given in Fig.~\ref{PDFR1}.
The contribution from Fig.~\ref{PDFR1}(c) is again identically zero due to $\bn^2=0$.  The contribution from  Fig.~\ref{PDFR1}(b) 
is nonsingular in the limit $x\rightarrow 1$. Its contribution should be omitted since it is subleading in the SCET expansion parameter, $\lambda$, which for DIS in
the threshold region has to be taken as $\lambda^2\cong 1-x$. The remaining contribution from  Fig.~\ref{PDFR1}(a) is given by:
\bea
{\hat \phi}^{(a)}(x)&=&2\alpha_s C_F(\mu^2)^\e \int \frac{d^d k}{(2\pi)^{d-2}}\frac{{\tilde p}-k^+}{k^+k^-}\delta(k^2)\delta(k^+-(1-x){\tilde p})\nonumber\\
&&=\frac{\alpha_s}{4\pi}C_F\left(\frac{\mu^2}{Q^2}\right)^\e x\left[-\frac{1}{\eir}\delta(1-x)+D_0(x)\right]\left(\frac{2}{\euv}-\frac{2}{\eir}\right)\,\,.
\label{pdfr}
\eea
Comparing the last result with the contribution from Fig.~\ref{SOFTR}(a) given in Eq.~(\ref{ssr1}) we see  that in the $x \to
1$ limit both contributions are identical. 

For the PDF the leading contribution in the $x\to 1$ limit (or equivalently in 
$\lambda$) comes entirely from the zero-bin region as we can see from the second $\delta $-function in Eq.~(\ref{pdfr}). $k^+$
which is supposed to be  $O(Q)$ (since $k$ is collinear to the incoming parton momentum with $p_1^+={\tilde p}=Q/{\sqrt 2}$)
is restricted to be equal to  $(1-x){\tilde p}$ which is $O(Q\lambda^2)$. Since  $k^-$ is also  $O(Q\lambda^2)$ then the first
$\delta$-function enforces the transverse components to be $O(Q\lambda^2)$. Thus the contribution comes
from the zero-bin (soft) region.  Notice that  Fig.~\ref{PDFR1}(a) is the only diagram relevant to the PDF as $x\rightarrow 1$
because we have defined our PDF with the naive collinear fields, which allowed us to integrate over all momentum space
including the zero-bin region. If we had restricted the SCET definition of the PDF to contain the purely collinear fields
(i.e., so $k^+$ is  $O(Q)$) then  Fig.~\ref{PDFR1}(a) would not contribute as the second $\delta$-function in Eq.~(\ref{pdfr})
could not be  satisfied by power counting arguments. Existing SCET treatments of DIS in the  $x \to 1$ limit differ in their
treatment of  Fig.~\ref{PDFR1}(a)~\cite{Becher:2006mr,Chay:2005rz}.  Ref.~\cite{Becher:2006mr} drops this diagram and argues
that collinear modes cannot contribute because  of kinematic constraints. This is essentially equivalent to our argument above
that the diagram receives support only from modes whose scaling is soft rather than  collinear. In our approach this
contribution is removed from the purely collinear jet function by a zero-bin subtraction. Our purely collinear PDF (i.e. the
naively  evaluated PDF minus the zero-bin subtraction) is analogous to the function $g_P$ that appears in the factorization
theorem of  Ref.~\cite{Chay:2005rz}. However, zero-bin subtractions are required for both the PDF and the jet function, as we will
see below, whereas Ref.~\cite{Chay:2005rz} only includes a zero-bin subtraction in their evaluation of the PDF. 

\begin{figure}[h]
\begin{center}
\includegraphics{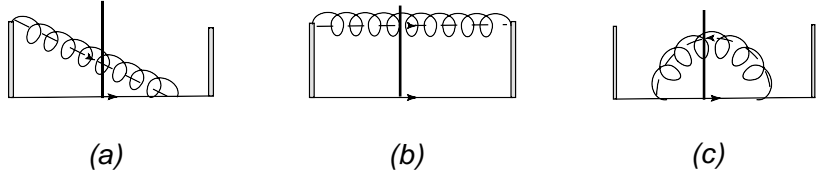}
\end{center}
\vskip -0.7cm \caption{Real gluon contribution to the outgoing jet
function.}
\label{JETR}
\end{figure}

The final state jet function  represents  outgoing (in the $\bar n$ direction) collinear partons with invariant mass  $Q^2(1-x)$, which is 
assumed to be much larger than $\Lambda^2_{\rm QCD}$. To define the jet function, we start with the following two-point correlation 
function  of collinear fields \cite{Bauer:2001yt}
   \begin{equation}
 \langle 0|T\left[ {\hat W}_{\bn}^\dagger(z){\hat \xi}_{\bn}(z)
  \bar {\hat \xi}_{\bn}(0) {\hat W}_{\bn}(0)\right]|0\rangle
   = i \frac{\not\! {\bn}}{\sqrt 2}\int \frac{d^4k}{(2\pi)^4}
   e^{-ikz} {\hat {\cal J}}(k) \,\, .
 \end{equation}
The naive dimensionless jet function, to be denoted by ${\hat J}(Q^2,x)$, is related to the absorptive part of ${\hat {\cal J}}$ and normalized at leading order to $\delta(1-x)$, 
so that  ${\hat J}(Q^2,x)=\frac{-1}{\pi}{\tilde p}\times\mathrm {Im}{\hat {\cal J}}$.
This definition is the gauge invariant version of the jet function defined in Ref.~\cite{Sterman:1986aj} with the 
full QCD fields replaced by the naive SCET ones. The Feynman diagrams with real gluon emission that contribute to the jet function are given in Fig.~\ref{JETR}.

The contribution of  Fig.~\ref{JETR}(c) is zero due to ${\bar n}^2=0$. From   Fig.~\ref{JETR}(a) we have
\begin{eqnarray}
\label{jet110} {\hat J}^{(a)}(Q^2,x)&=&\alpha_s C_F(\mu^2)^\e \int \frac{d^d
k}{(2\pi)^{d-2}}\frac{8{\tilde p}l^-}{k^-l^2}(l^- -k^-)\delta(k^2)
\delta((l-k)^2)
\end{eqnarray}
where $l$ is the  momentum for the outgoing jet, whose components are $l^+=\frac{Q}{\sqrt 2}(1-x)$; $l^-=\frac{Q}{\sqrt 2}$ and $l_{\perp}=0$. $k$ is collinear to the
outgoing quark momentum with $k^- \cong Q$, $k^+ \cong Q\lambda^2$ and $k_\perp \cong Q\lambda$.  Recalling that in threshold region for DIS we identify $1-x\cong
\lambda^2$ then simple power counting shows that the above contribution scales as $1/\lambda^2$. 
Carrying out the integrations over $k^+$, $k_\perp$, and $k^-$ (in that order) we find
\bea
\label{jet2} {\hat J}^{(a)}(Q^2,x) &=&\frac{\alpha_s}{4\pi}C_F\times 4\left[\left(\frac{1}{\eir^2}+\frac{1-\lnQ}{\eir}\right)\delta(1-x)-\frac{D_0(x)}{\eir}-D_0(x)+D_1(x)\right.\nonumber\\
&&\left.+\left(\frac{1}{2}\llnQ-\lnQ+2-\frac{\pi^2}{4}\right)\dd+\lnQ D_0(x)\right]\,\,,
\eea

The contribution from  Fig.~\ref{JETR}(c) can be obtained in similar manner and is given by
\begin{eqnarray}
 {\hat J}^{(c)}(Q^2,x) &=&\frac{\alpha_s}{4\pi}C_F\left[-\frac{\delta(1-x)}{\eir}-\left(1-\lnQ \right)\delta(1-x)+D_0(x) \right]\,\, ,
\end{eqnarray}
Including the contribution from the mirror of  Fig.~\ref{JETR}(a) we get for the real gluon contribution to the jet function
\bea
{\hat J}_{\rm real}(Q^2,x)&=&\frac{\alpha_s}{4\pi}C_F\left[\left(\frac{4}{\e_{\rm
IR}^2}+\frac{3-4\ln \left(\frac{Q^2}{\mu^2}\right)}{\e_{\rm
IR}}\right)\delta(1-x)-\frac{4D_0(x)}{\e_{\rm
IR}}+4D_1(x)\right.\\
&&\left.+4\ln \left(\frac{Q^2}{\mu^2}\right)D_0(x)+\left(2\ln^2
\left(\frac{Q^2}{\mu^2}\right)-3\ln
\left(\frac{Q^2}{\mu^2}\right)+7-\pi^2\right)\delta(1-x)\right]\,\, . \nonumber
\label{jetr}
\eea
Let us now consider the zero-bin contribution included in the above result. For Fig.~\ref{JETR}(a) we take the gluon momentum $k$ to be soft. 
Then we ignore $k^-$ relative $l^-$ in the numerator of Eq.~(\ref{jet110}) and we drop $k^2$ in the last $\delta$-function as it scales as $\lambda^4$ compared with $l^2 \cong Q^2\lambda^2$. 
We then get a contribution that also scales as $1/\lambda^2$ like in the collinear region:
\bea
J_{\rm zb}^{(a)}(Q^2,x) &=&2\alpha_s C_F (\mu^2)^{\e}\frac{1}{1-x}
\int \frac{d^d k
}{(2\pi)^{d-2}}\frac{1}{k^-} \delta (k^2)\delta(k^+-l^+)\nonumber\\
&&=\frac{\alpha_s}{4\pi}C_F\left(\frac{\mu^2}{Q^2}\right)^\e\left[-\frac{1}{\eir}\delta(1-x)+D_0(x)\right]\left(\frac{2}{\euv}-\frac{2}{\eir}\right)\,\,,
\label{zbr}
\eea
The last result is exactly the same as the ones given in Eq.~(\ref{ssr1}) and
Eq.~(\ref{pdfr}) (when taken to the $x \to 1$ limit.) Noticing that the zero-bin contribution from Fig.~\ref{JETR}(b) is subleading
in $\lambda$ so the soft contribution to the real gluon emission for the jet function comes only  from Fig.~\ref{JETR}(a).

We now include the virtual contributions to the soft factor, PDF and jet function. For the soft factor the only contribution comes from Fig.~\ref{fullcurrent}(c).
The sum of the real and virtual contributions is
\bea
S(1-x)=\delta(1-x)+\asf\times4\left(\frac{1}{\euv}-\frac{1}{\eir}\right)\left[-\frac{1}{\euv}\delta(1-x)+D_0(x)\right]\,\,.
\eea
Note that the soft function is scaleless and the final answer is proportional to $\frac{1}{\euv}-\frac{1}{\eir}$.
The factor $(\mu^2/Q^2)^\epsilon$ can be set to unity since when we expand this factor in powers of $\epsilon$ and mulitply
by $\frac{1}{\euv}-\frac{1}{\eir}$, the finite logarithms cancel. This makes physical sense since this quantity is scaleless.
A similar situation arises for the PDF, $\hat{\phi}(x)$, however, the jet function will have logarithms of $Q$ because it knows about the
scale $Q\sqrt{1-x}$. The renormalized soft function to $O(\alpha_s)$ is  
\bea
S^R(1-x)=\delta(1-x)+\asf \times(-4)\frac{D_0(x)}{\eir}\,\, .
\eea
The large $N$ moments of the soft function are 
\bea
S^R_N=1+\asf \times 4\frac{\ln {\overline N}}{\eir}\,\,,
\eea
with $\ol=Ne^{\gamma_E}$. For the PDF, after 
 including virtual contributions, we find
\bea
{\hat \phi}(x)=\delta(1-x)+\asf\left[3\left(\frac{1}{\euv}-\frac{1}{\eir}\right)\delta(1-x)+4D_0(x)\left(\frac{1}{\euv}-\frac{1}{\eir}\right)\right]\,\,.
\label{ff11}
\eea
The renormalized PDF is
\bea
{\hat \phi}^R(x)=\delta(1-x)+\asf\left(-\frac{1}{\eir}\right)\left[3\delta(1-x)+4D_0(x)\right]\,\,,
\eea
and in moment space we get
\bea
{\hat \phi}^R_N=1+\asf\left(-\frac{1}{\eir}\right)\left[3-4\ol \right]\,\,,
\eea
which is a well-known result. The anomalous dimension of the (naive) PDF can be immediately read off from the UV poles in Eq.~(\ref{ff11}) :
\bea
\gamma_{2,N}=\asf \times2[4\ol-3]\,\,.
\eea

The virtual contribution to the jet function comes from Fig.~\ref{fullcurrent}(b), its mirror diagram, and the wave function
renormalization. In the result for $I_{\bar n}$ (which is equal to $I_n$ given in Eq.~(\ref{In})) we expand the factor $(\mu^2/Q^2)^\e$
and  get for the naive jet function
\bea\label{njf}
{\hat J}(Q^2,x)&=&\delta(1-x)+\asf \left[-\frac{4D_0(x)}{\eir}+4D_1(x)+4\lnQ D_0(x)\right.\nonumber\\
&&\left.+\left(2\llnQ-3\lnQ +7-\pi^2\right)\delta(1-x)\right]\nonumber\\
&&+\asf\left(\frac{\mu^2}{Q^2}\right)^\e  \left[\frac{4}{\euv\eir}+\frac{3}{\euv}\right]\delta(1-x)\,\,.
\eea
Next we have to include the zero-bin subtraction which as we have seen earlier is equivalent to subtracting the one-loop soft contribution. It is easy to see the 
effect of the zero-bin subtraction is to replace  $1/\epsilon_{\rm IR}$ everywhere in Eq.~(\ref{njf}) with $1/\epsilon_{\rm UV}$. 
The remaining UV divergences can now be removed by counterterms and  the renormalized jet function is
\bea
{J}^R(Q^2,x)&=&\delta(1-x)+\asf \left[ 4D_1(x)+4\lnQ D_0(x)\right.\nonumber\\
&&\left.+\left(2\llnQ-3\lnQ +7-\pi^2\right)\delta(1-x)\right]\,\,.
\eea

The results presented so far for the zero-bin contributions (real and virtual) for the naive PDF and jet function show that the renormalized soft factor 
has to be subtracted from each one of these functions to obtain the truly collinear contributions. Thus we find that the renormalized collinear matrix elements are
\bea
\phi^R_N=\left(\frac{{\hat \phi}_N}{S_N}\right)^R\,\,\,\,~~~~~ J^R_N=\left(\frac{{\hat j}_N}{S_N}\right)^R\,\,,
\label{cor}
\eea
which to $O(\alpha_s)$ are given by
\bea
\phi^R_N=1+\asf\times(-3)\frac{1}{\eir}\,\,,
\eea
and 
\bea
J_N^R=1+\asf \left[2\ln^2\left(\frac{Q^2}{\overline{N} \mu^2}\right)
-3\ln\left(\frac{Q^2}{ \overline{N} \mu^2}\right)+7-\frac{2}{3}\pi^2\right]\,\,.
\label{jr}
\eea
The last result for the collinear jet is finite and is equal to the matching coefficient for DIS at the intermediate scale,
$\mu^2 =Q^2/{\overline N}$, in the analysis of DIS as $x \to 1$ in Ref.~\cite{Manohar:2003vb} .
Thus the factorization theorem for DIS in the threshold region reads
\bea
F_{2,N}=H(Q^2/\mu^2)\, J^R_N\, \phi^R_N\, S^R_N=H(Q^2/\mu^2)\left(\frac{{\hat J}_N}{S_N}\right)^R\left(\frac{{\hat \phi}_N}{S_N}\right)^R S^R_N\,\,.
\label{fun11}
\eea
where $H(Q^2/\mu^2)$ is the square of the matching coefficient $C(Q^2/\mu^2)$ given in Eq.~(\ref{cc}). With the above results for 
$J^R_N$, $\phi^R_N$ and $S^R_N$ we get
\bea
H(Q^2/\mu^2)\, J^R_N\, \phi^R_N\, S^R_N &=&1+\asf \left \{-\frac{1}{\eir}[3-4\ol]-2\llnQ+6\lnQ \right.\nonumber\\
&&\left.+2\ln^2 \left(\frac{Q^2}{{\overline N}\mu^2}\right)-3\ln \left(\frac{Q^2}{{\overline N}\mu^2}\right)-9-\frac{\pi^2}{3}\right \}\,\,,
\eea
which agrees with the moments of DIS structure function in the large-$N$ limit calculated in full QCD. It is straightforward to show that the results
in Eq.~(\ref{cor}) which were shown to hold to ${\cal O}(\alpha_s)$ in pure DR can be obtained by performing field
redefinitions \cite{Lee:2006nr} on the naive SCET fields and one obtains the naive collinear matrix elements divided by the 
soft Wilson line matrix elements as in Eq.~(\ref{dc}).

Based on the factorized form of the non-singlet DIS structure function given in Eq.~(\ref{fun11}) we now comment on the resummation of the large logarithms in the
threshold region for DIS. In moment space  Eq.~(\ref{fun11}) can be written in the following form (henceforth we drop the superscript $R$ with the understanding 
that we consider only renormalized quantities)
\bea
F_{2,N}=H(Q^2/\mu^2) J_N\left(\frac{Q^2}{{\overline N}\mu^2}\right) \hat\phi_N(\mu^2)\,\,.
\label{fun2}
\eea
The hard part $H$ depends only on $Q^2/\mu^2$ and it is obtained by matching the full QCD current onto the SCET one
(at the higher scale $Q^2$) order by order in perturbation theory. The anomalous dimension of the SCET current
$\gamma_1$ is then calculated from the matching coefficient (through Eq.~(\ref{ano})) and is used to run down to the
intermediate scale of DIS: $Q^2/{\overline N}$. The quantity $J_N$, which is IR safe, depends only on the
intermediate scale  as can be seen   in Eq.(\ref{jr}). Thus the first two terms on the right hand side of Eq.~(\ref{fun2}) are perturbative and IR safe.
Below the intermediate scale we are left with only one non-perturbative quantity, the PDF taken to the large $N$ limt. 
By exploiting  the standard Altarelli-Parisi kernels with anomalous dimension $\gamma_2$ (taken to the large $N$-limit) we can then
evolve the PDF to an arbitrary factorization scale. The two stage running between $Q^2$ and $Q^2/{\overline N}$ with
$2 \gamma_1$ and between $Q^2/{\overline N}$ and some arbitrary factorization scale $\mu_F$ with $\gamma_2$ resums all
the large logarithms in moment space. This has been established to be equivalent to the standard pQCD
resummation~\cite{Sterman:1986aj,Catani:1989ne} in Ref.~\cite{Idilbi:2006dg} to all  orders in the  strong coupling
constant and to arbitrary sub-leading logarithms.

\section{Zero-Bin At Higher Orders}

In this section we consider the abelian virtual two-loop diagrams that contribute
to the $n$-collinear jet function $\langle 0 \vert W^{(0)\dagger}_n \xi^{(0)}_n \vert q(p)\rangle$ 
that appears in the factorization theorem for the quark form factor.
We will show how dividing by the soft factor reproduces the zero-bin subtraction for these diagrams. 

\begin{figure}[!h]
 \begin{center}
 \hspace{-0.5 in} \includegraphics[width=4.5in]{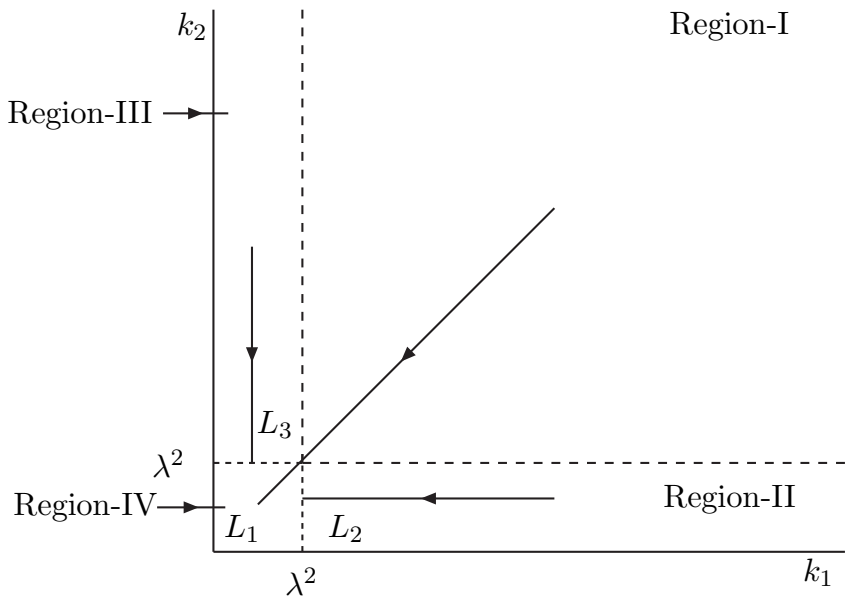}
 \end{center}
\caption{Zero-Bin Regions At Two-Loops.}\label{zbtl}
 \end{figure}

\begin{figure}[!h]
 \begin{center}
 \includegraphics{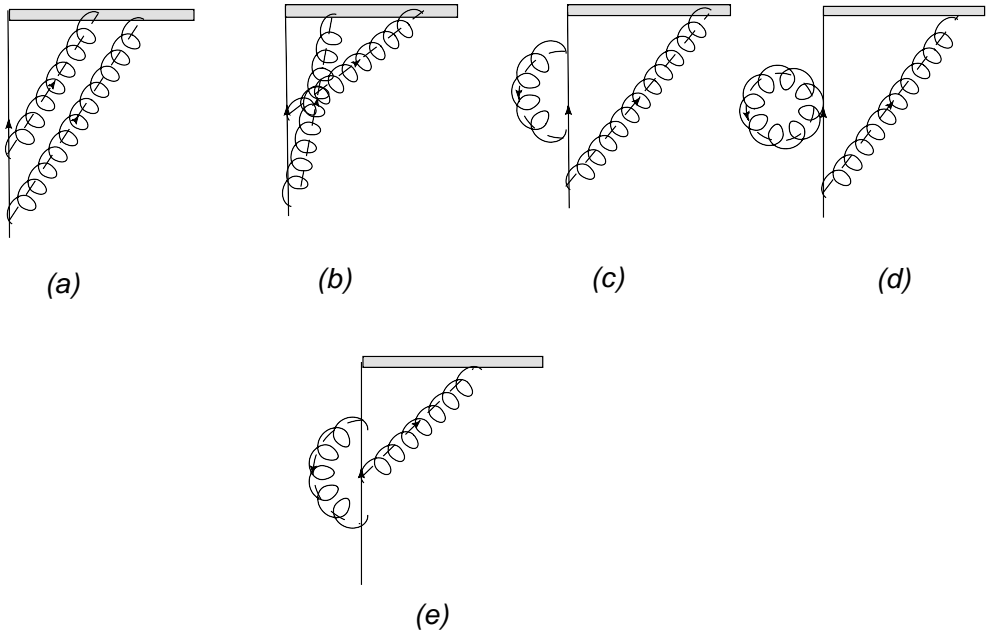}
 \end{center}
 \vskip -0.2cm \caption{All the two-loop abelian diagrams that contribute the $n$-collinear jet function.}
  \label{twoloop}
 \end{figure}

The abelian two-loop diagrams are shown in Fig.~\ref{twoloop}. Before considering these in detail 
we make some general comments about the zero-bin subtraction at two loops. Let
$k_1$  and $k_2$ be the loop momenta, which are routed so that $k_1$ and $k_2$ correspond to the 
virtual gluon momenta, since the zero-bin associated with any fermion lines is easily checked to 
be subleading in the $\lambda$ expansion.  Since we integrate over all momentum space, we need to 
subtract the contribution where either of the $k_i$ scales like a soft momentum rather than collinear. 
In Fig.~\ref{zbtl}, the region where $k_1$ ($k_2$) is collinear is separated
from the region where $k_1$ ($k_2$) is soft by the vertical (horizontal) dotted line.
The collinear matrix element gets contributions only from Region-I. Region-II
corresponds to $k_2$  soft and $k_1$ collinear, Region-III corresponds to 
$k_1$  soft and $k_2$ collinear, and Region-IV corresponds to both $k_1$ and $k_2$ soft.

Denote the integrand of the two-loop integral ${\cal I}(k^c_1,k^c_2)$, where the superscript $c$
in $k_i^c$ means that $k_i$ is assumed to have collinear scaling. The naive two-loop integral
is
\bea
(\mu^2)^{2 \epsilon}\int \frac{d^d k_1}{(2\pi)^d}\int \frac{d^d k_2}{(2\pi)^d}{\cal I} (k^c_1,k^c_2) \equiv 
\int_{k_1,k_2} {\cal I} (k^c_1,k^c_2) \, .
\eea 
The integrands for the zero-bin subtractions for Region-II and Region-III are $-{\cal I}(k^c_1,k^s_2)$ and  
$-{\cal I}(k^s_1,k^c_2)$, respectively, where the superscript $s$ in $k_i^s$ means that the integrand is evaluated 
assuming that $k_i$ satisfies the soft scaling and the integrand is expanded to lowest order in $\lambda$.
 The zero-bin subtraction from Region-IV  is subtle 
because this Region has been doubly counted both in the original naive collinear integrals and 
also in the zero-bin subtraction for Region-II and  Region-III. From the original integral we need to subtract a 
contribution in which both $k_1$ and $k_2$ are soft. We will denote the integrand for this 
zero-bin as ${\cal I}_{L_1}(k^s_1,k^s_2)$, where $L_1$ denotes the limit when $k_1$ and $k_2$ are taken to be 
soft simultaneously.  From the zero-bin subtraction for Region-II,
we must perform a second subtraction that comes from $k_1^c$ becoming soft after having first made the 
soft approximation for $k_2$. We call the integrand for the zero-bin subtraction defined by this limit
${\cal I}_{L_2}(k_1^s,k_2^s)$.
Likewise, we have a similar subtraction from the zero-bin of Region-III, denoted ${\cal I}_{L_3}(k_1^s,k_2^s)$.
In general, the integrand ${\cal I}_{L_i}(k_1^s,k_2^s)$ depends on the order  in which we take the soft limits, so there are really
five different zero-bin subtractions in the two-loop calculation. The result for the collinear contribution to each 
two-loop diagram is
\bea\label{mastzb}
\int_{k_1,k_2} \left( {\cal I}(k_1^c,k_2^c) - \bigg[ {\cal I}(k_1^c,k_2^s) -{\cal I}_{L_2}(k_1^s,k_2^s) \bigg]
-   \bigg[{\cal I}(k_1^s,k_2^c) -{\cal I}_{L_3}(k_1^s,k_2^s)  \bigg] - {\cal I}_{L_1}(k_1^s,k_2^s) \right) \, .
\eea
The various order of limits defining the integrands ${\cal I}_{L_i}(k_1^s,k_2^s)$ are shown in Fig.~\ref{zbtl}.


Next we turn to the evaluation of the individual diagrams.  Figs.~\ref{twoloop}(a)
\ref{twoloop}(c) and \ref{twoloop}(d) have color factors proportional to $C_F^2$ while the color factors
for   Figs.~\ref{twoloop}(b) and \ref{twoloop}(e) are $C_F(C_F-C_A/2)$. The results
for an abelian theory are obtained by taking $C_F \to 1$ and $C_A \to 0$, in which case the color factor
is one for all graphs in Fig.~\ref{twoloop}.

Consider first the two-loop SCET diagrams in Fig.~\ref{twoloop}(a) and 
\ref{twoloop}(b). 
Working in Feynman gauge, using pure DR as a regulator, and ignoring the
$i\epsilon$'s  in the propagators as they are not relevant to
our discussion, we get from Fig.~\ref{twoloop}(a),
\begin{eqnarray}
\label{twog1}
I^{(a)}=-2g_s^4 \int_{k_1,k_2}\frac{(p^+-k_1^+-k_2^+)}{(p-k_1-k_2)^2k_1^2k_2^2(k_1^++k_2^+)}
\left[\frac{(p^+-k_1^+)}{(p-k_1)^2 k_1^+}+\frac{(p^+-k_2^+)}{(p-k_2)^2k_2^+}\right] \ ,
\end{eqnarray}
where $k_i$ are both collinear to the incoming parton with momentum $p$. 
From Fig.~\ref{twoloop}(b) we have
\begin{eqnarray}
\label{twog2}
I^{(b)}=-2g_s^4 \int_{k_1,k_2} \frac{(p^+-k_1^+-k_2^+)}{(p-k_1-k_2)^2k_1^2k_2^2(k_1^++k_2^+)}
\left[\frac{(p^+-k_1^+)}{(p-k_1)^2 k_2^+}+\frac{(p^+-k_2^+)}{(p-k_2)^2 k_1^+}\right] \ ,
\end{eqnarray}
In Eqs.~(\ref{twog1}) and (\ref{twog2}), we have chosen to write the integrand so it is symmetric under interchange of $k_1$ and $k_2$.

The integrands for  Figs.~\ref{twoloop}(a) and \ref{twoloop}(b) in Region-IV are
\begin{eqnarray}
\label{zb11}
 {\cal I}_{L_1}^{(a)}(k_1^s,k_2^s)=-g_s^4
\frac{1}{2}\frac{1}{k_1^2k_2^2(k_1^-+k_2^-)(k_1^++k_2^+)}\left[\frac{1}{k_1^-k_1^+}+\frac{1}{k_2^-k_2^+}
\right]\,\, ,
\end{eqnarray}
and
\begin{eqnarray}
\label{zb12}
 {\cal I}_{L_1}^{(b)}(k_1^s,k_2^s)=-g_s^4C_F^2\times \frac{1}{2}\frac{1}{k_1^2k_2^2(k_1^-+k_2^-)(k_1^++k_2^+)}
 \left[\frac{1}{k_1^-k_2^+}+\frac{1}{k_2^-k_1^+}\right]\,\, .
\end{eqnarray}
Note that the contributions in Eq.~(\ref{zb11}) and Eq.~(\ref{zb12}) are not subleading in SCET power counting. For 
the two loop momenta in the soft region, the measure of the loop integral scales as $\lambda^{16}$ and 
each integrand scales as $1/{\lambda^{16}}$ thus the contribution is ${\cal O}(1)$.  
The sum of these two zero-bin contributions is 
\begin{eqnarray}
\int_{k_1,k_2} \left( {\cal I}_{L_1}^{(a)}(k^s_1,k^s_2)+{\cal I}_{L_1}^{(b)}(k^s_1,k^s_2) \right) 
=-g_s^4 \int_{k_1,k_2}  \frac{1}{2}\frac{1}{(k_1^2k_1^-k_1^+)(k_2^2k_2^-k_2^+)} =\frac{1}{2}I^2_s\,\, ,
\end{eqnarray}
where $I_s$ is the one-loop soft contribution given in Eq.~(\ref{is}). The  two-loop result is consistent with 
the exponentiation theorem for the abelian amplitudes of soft gluon radiation
in a web (i.e., two soft Wilson lines).  

Now we consider the case when one collinear momentum becomes soft while the other is kept collinear.
 In Eq.~(\ref{zb11}) let us take $k_1$ to the zero-bin region. The integrand is then
\begin{eqnarray}
 {\cal I}^{(a)}(k_1^s,k^c_2)=-2g_s^4\frac{(p^+-k_2^+)}{[(p-k_2)^2-2k_1^-(p^+-k_2^+)]k_1^2k_2^2k_2^+}
 \left[\frac{p^+-k_2^+}{(p-k_2)^2k_2^+}+\frac{1}{-2k_1^-k_1^+}\right]\,\, .
\end{eqnarray}
The measure in the loop integral in this case scales as $\lambda^{12}$ because there is one collinear and
one soft loop momentum. The factor outside the brackets in the integrand scales as $\lambda^{-8}$, the
first factor inside square brackets scales as $\lambda^{-2}$ and the second scales as $\lambda^{-4}$.
Therefore, only the second term contributes so we  get
\begin{eqnarray}
 {\cal I}^{(a)}(k_1^s,k_2^c)=-2g_s^4 \frac{(p^+-k_2^+)}{[(p-k_2)^2-2k_1^-(p^+-k_2^+)]k_1^2k_2^2k_2^+}
 \left[ \frac{1}{-2k_1^-k_1^+}\right]\,\, .
\end{eqnarray}
Similarly for $k_1$ collinear and $k_2$ soft we find
\begin{eqnarray}
 {\cal I}^{(a)}(k_1^c,k_2^s)=-2g_s^4 \frac{(p^+-k_1^+)}{[(p-k_1)^2-2k_2^-(p^+-k_1^+)]k_1^2k_2^2k_1^+}
 \left[\frac{1}{-2 k_2^- k_2^+}\right]\,\, .
\end{eqnarray}
The analogous zero-bins for Fig.~\ref{twoloop}(b) are
\begin{eqnarray}
 {\cal I}^{(b)}(k_1^s,k_2^c)=-2g_s^4 \frac{(p^+-k_2^+)}{[(p-k_2)^2-2k_1^-(p^+-k_2^+)]k_1^2k_2^2k_2^+}
 \left[\frac{p^+-k_2^+}{(p-k_2)^2k_1^+}\right]\,\, ,
\end{eqnarray}
and
\begin{eqnarray}
{\cal I}^{(b)}(k_1^c,k_2^s)=-2g_s^4 \frac{(p^+-k_1^+)}{[(p-k_1)^2-2k_2^-(p^+-k_1^+)]k_1^2k_2^2k_1^+}
\left[\frac{p^+-k_1^+}{(p-k_1)^2k_2^+}\right]\,\, .
\end{eqnarray}
Upon summing  the zero-bins for each Region, the integrand factorizes. For Region-II, we get
\bea\label{RII}
{\cal I}^{(a)}(k_1^c,k_2^s) + {\cal I}^{(b)}(k_1^c,k_2^s) =
g_s^4\,\frac{(p^+-k_1^+)}{(p-k_1)^2k_1^2k_1^+}\times \frac{1}{k_2^2k_2^-k_2^+}
\eea
and for Region-III we obtain the same with $k_1$ and $k_2$ interchanged. The zero-bins
for Region-II and Region-III combine to give 
\begin{eqnarray}
\label{ns}
\int_{k_1,k_2} \left( {\cal I}^{(a)}(k_1^s,k_2^c)+{\cal I}^{(a)}(k_1^c,k_2^s)+{\cal I}^{(b)}(k_1^s,k_2^c)+{\cal I}^{(b)}(k_2^s,k_1^c) 
\right) = I_n\cdot  I_s\,\, ,
\end{eqnarray}
where $I_n$ is the one-loop contribution to the $n$-collinear jet in Eq.~(\ref{coln1}). 
Finally, we have to perform the second zero-bin subtraction from the zero-bins corresponding to Region-II and Region-III,
i.e.  the  terms with integrands ${\cal I}_{L_2}(k_1^s,k_2^s)$ and
${\cal I}_{L_3}(k_1^s,k_2^s)$ in Eq.~(\ref{mastzb}). It is clear from Eq.~(\ref{RII}) that this is simply $\frac{1}{2}I_s^2$ for 
each Region. Thus the final result for Figs.~\ref{twoloop}(a) and \ref{twoloop}(b)
including all zero-bin subtractions is
\begin{eqnarray}\label{tlzbs}
I^{(a)}+I^{(b)}-[I_n\cdot I_s- I_s^2]-\frac{1}{2}I_s^2 
=I^{(a)}+I^{(b)}-I_n\cdot I_s+\frac{1}{2}I_s^2 \,\, ,
\end{eqnarray}
where $I^{(a)}$ and $I^{(b)}$ correspond to the naive evaluation of Figs.~\ref{twoloop}(a) and \ref{twoloop}(b), respectively.

Next we turn to the evaluation of Figs.~\ref{twoloop}(c)-(e). Denote the momentum 
that flows into the gluon attached to the collinear Wilson line as $k_2$ while the momentum flowing through the
other gluon as $k_1$. For these three diagrams it is easy to show that the only zero-bin contribution
which is leading in $\lambda$ comes from the region where $k_1$ is collinear  and $k_2$ is soft.
In Figs.~\ref{twoloop}(c) and (d), $k_1$ flows through the self-energy subgraph.
The self-energy in the collinear sector of SCET is evaluated in Ref.~\cite{Bauer:2001yt} with the
result that 
\bea\label{se}
\Sigma(q^2) = \frac{\alpha_s}{4\pi}\frac{q^2}{2 q^+} \frac{\nsl}{2} \frac{\Gamma(\epsilon)\Gamma(1-\epsilon)\Gamma(2-\epsilon)}{\Gamma(2-2\epsilon)}
\left(\frac{-q^2}{e^{\gamma_E} \mu^2}\right)^{-\epsilon} \, ,
\eea
where $q^2$ is the virtuality, which in our case is $q^2 =(p-k_2)^2 = -2 p^+ k_2^-+{\cal O}(\lambda^4)$. Inserting this result into the two-loop graphs
in Figs.~\ref{twoloop}(c) and (d) yields the integral
\bea
&&\int_{k_1,k_2} \left( {\cal I}^{(c)}(k_1^c,k_2^s)+{\cal I}^{(d)}(k_1^c,k_2^s) \right) \\ && 
\qquad =  i \alpha_s^2\frac{\Gamma(\epsilon)\Gamma(1-\epsilon)\Gamma(2-\epsilon)}{\Gamma(2-2\epsilon)} ({e^{\gamma_E} \mu^4})^\epsilon
\int_{k_2}\frac{1}{k_2^+ k_2^- k_2^2} (2 p^+k_2^-)^{-\epsilon} \, .\nn
\eea
To evaluate  Fig.~\ref{twoloop}(e), we perform the $k_1$ integral which is simply the collinear vertex correction in the limit 
that the external gluon is taken to be soft. Equivalently, this subgraph is simply the one-loop vertex correction to the 
coupling between the collinear quark and soft gluon. The result after doing this integral is
\bea
\int_{k_1,k_2} {\cal I}^{(e)}(k_1^c,k_2^s) =
-i \alpha_s^2\frac{\Gamma(\epsilon)\Gamma(1-\epsilon)\Gamma(2-\epsilon)}{\Gamma(2-2\epsilon)} ({ e^{\gamma_E} \mu^4})^\epsilon
\int_{k_2}\frac{1}{k_2^+ k_2^- k_2^2} (2 p^+k_2^-)^{-\epsilon} \nn
\eea
We find that the zero-bins for Fig.~\ref{twoloop}(c),(d) and (e) add up to zero. The cancellation can be partly
understood on the basis of the QED Ward identity, which relates the UV divergent pieces in the self-energy and vertex correction.

There are no zero-bin subtractions beyond those given in Eq.~(\ref{tlzbs}). We now compare this result with the soft subtraction, Eq.~(\ref{fun}), which 
yields
\bea
\frac{\langle 0 |\hat{W}^{\dagger}_n {\hat \xi}_n | q(p_1)\rangle}{\langle 0 |Y^{\dagger}_{\nb} Y_n | 0\rangle}
&=& \frac{1 + I_n +I^{(a)-(e)} + ...}{1 + I_s + \frac{1}{2}I_s^2 + ...} \\
&=& 1 +(I_n-I_s) + \left(I^{(a)-(e)}-I_n\cdot I_s + \frac{1}{2}I_s^2 \right) + ... \, . \nn
\eea
In the right hand side of the first line, we have expanded the numerator and denominator to $O(\alpha_s^2)$, separately. The
one-loop naive collinear integral is $I_n$, $I^{(a)-(e)}$ is the naive collinear  evaluation of the sum of graphs in
Fig.~\ref{twoloop}, $I_s$ is the one-loop soft integral, and the second order term in the denominator is a consequence of the
exponentiation theorem of the abelian contributions inside a web mentioned earlier. In the second line, we have expanded the
quotient to second order in $\alpha_s$. The  $O(\alpha_s^2)$ term is consistent with Eq.~(\ref{tlzbs}) and the cancellation
of the zero-bin in the sum of Figs.\ref{twoloop}(c),(d) and (e). Therefore, dividing by the soft Wilson line and the zero-bin
subtraction give the same result. It would be interesting to extend the analysis of this section to the non-abelian theory.

\section{Conclusions}

In this paper we have considered the two prescriptions deivsed to remove overlapping contributions to collinear matrix elements,  the soft and
zero-bin subtractions. We have demonstrated explicitly the equivalence of the two prescriptions for the abelian contributions to the quark form
factor up to two-loop level using DR to regularize both UV and IR  divergences. We also studied  DIS in the threshold region to ${\cal
O}(\alpha_s)$ in SCET. In our treatment for DIS all Wilson lines were defined on the light-cone.The essential result is that soft contributions
to naively defined collinear matrix elements have to be subtracted  in order to derive  proper factorization theorems  as in Eq.~(\ref{dc}). We
have shown by explicit calculation that soft and zero-bin subtractions  are equivalent in the examples studied in this paper.  Our results
obtained by fixed order pertubative calculations can be extended to all orders in the strong coupling by performing  field redefinitions as
proposed in Ref.~\cite{Lee:2006nr} with a suitable choice of IR regulators.

\acknowledgments 
 This work was supported in part by the Department
of Energy under grant numbers DE-FG02-05ER41368, DE-FG02-05ER41376, and DE-AC05-84ER40150.
Partial support of the U.S. Department of Energy under grant no. DE-FG02-93ER-40762 is
acknowledged.
We thank G.~Sterman, X.~d.~Ji and S.~Fleming for useful discussions. 
One of us (A.I.) presented results in Sections II and III   at the ECT$^*$
``Heavy Quarkonium and Related Heavy Quark States''  workshop,  August 28, 2006. The talk 
is available on the web at http://www.phy.duke.edu/\~{}mehen/ECT/talks.html.

\section{Appendix}

Here we explain in detail why the poles in Eq.~(\ref{I4}) are UV 
using contour integration.
 Let us first integrate over the
light-cone component $k^-$ with contour integration. The
poles are
\begin{eqnarray}
k^-=\frac{\vert \vec {k}_\perp \vert^2}{2(k^+-{\tilde
p})}-\frac{i0}{2(k^+-{\tilde p})},~~~~k^-=\frac{\vert \vec {k}_\perp
\vert^2+2{\tilde p}k^+}{2k^+}-\frac{i0}{2k^+},~~~~k^-=i0.
\label{poles}
\end{eqnarray}
There are two regions that contribute, $0\leq k^+ \leq {\tilde p}$
and $k^+ \geq {\tilde p}$. For $0\leq k^+ \leq {\tilde p}$ 
we have to pick the second pole. For $k^+ \geq {\tilde p}$ we
have to pick the third pole. After performing the contour
integration we integrate over $\vert \vec {k}_\perp \vert^2$ in
$d=2-2\e$. This will introduce the first $\Gamma(\e_{\rm UV})$. We
then integrate over $k^+$. For the contribution from $0\leq k^+ \leq
{\tilde p}$, the integration over $k^+$ will introduce another pole
as $k^+ \rightarrow 0$,  $\Gamma(-\e)$. This pole has to be taken as
UV  for the following reason: When $k^+ \rightarrow 0$, then from
the second pole in Eq.~(\ref{poles}) we see that $k^- \rightarrow
\infty$. This combination clearly means that both  $k^0$ and $k^3$
approach infinity (in addition to the UV pole from $\vert \vec
{k}_\perp \vert^2\rightarrow \infty$). The same reasoning applies
for the contribution from the region $k^+ \geq {\tilde p}$ where we
get $\Gamma(-\e)$ from $k^+ \rightarrow \infty$. The third
pole in Eq.~(\ref{poles}) indicates $k^- \rightarrow 0$ so we
again have  one light-cone component approaching zero
while the other one approaches infinity. Thus both contributions
have a double UV poles and the total result is given in
Eq.~(\ref{I4}). The observation that a vanishing light-cone component may
lead to a UV divergence (as opposed to IR divergence) was also discussed in
Ref.~\cite{Bauer:2003td}.

\end{document}